\documentclass[12pt,a4paper]{article}

\usepackage{latexsym}
\usepackage{amsmath}
\usepackage{amsfonts}
\usepackage{amssymb}
\usepackage{amsthm}
\usepackage{amscd}
\usepackage{mathrsfs}

\newcommand{\be}{\begin{equation}}
\newcommand{\ee}{\end{equation}}
\newcommand{\bea}{\begin{eqnarray}}
\newcommand{\eea}{\end{eqnarray}}

\def\ad{{\mathrm{ad}}}                   %
\def\cO{{\cal O}}                        %
\def\bC{{\mathbb C}}                     %
\def\bN{{\mathbb N}}                     %
\def\1{{\mbox{\boldmath $1$}}}           %
\def\red{\mathrm{red}}                   %
\def\cC{{\mathcal C}}                    %
\def\cM{{\mathcal M}}                    %
\def\bR{{\mathbb R}}                     %
\def\bZ{{\mathbb Z}}                     %
\def\ri{{\mathrm{i}}}                    %
\def\tr{\mathrm{tr}}                     %
\def\diag{\mathrm{diag}}                 %
\def\cA{{\mathcal A}}                    %
\def\cG{{\mathcal G}}                    %
\newcommand{\oper}{\operatorname}        %
\begin{document}

\vspace*{0.5cm}
\medskip

\begin{center}
{\Large \bf An integrable $BC(n)$ Sutherland model with two types of particles}
\end{center}

\vspace{0.2cm}

\begin{center}
V.~Ayadi${}^{a}$  and L. Feh\'er${}^{a,b}$ \\

\bigskip

${}^a$Department of Theoretical Physics, University of Szeged\\
Tisza Lajos krt 84-86, H-6720 Szeged, Hungary\\
e-mail: ayadi.viktor@stud.u-szeged.hu

\bigskip

${}^{b}$Department of Theoretical Physics, MTA  KFKI RMKI\\
H-1525 Budapest, P.O.B. 49,  Hungary \\
e-mail: lfeher@rmki.kfki.hu

\bigskip

\end{center}

\vspace{0.2cm}

\begin{abstract}
A hyperbolic $BC(n)$ Sutherland model involving three independent coupling constants
that characterize the interactions of two types of particles
moving on the half-line is derived by Hamiltonian reduction of
the free geodesic motion on the group $SU(n,n)$.
The symmetry group underlying the reduction is provided by the direct product of
the fixed point subgroups of two commuting involutions of $SU(n,n)$.
The derivation
implies the integrability of the model and yields a
simple algorithm for constructing its solutions.

\end{abstract}

\newpage

\section{Introduction}
\setcounter{equation}{0}

The integrable many-body systems discovered by Calogero and Sutherland around 40 years ago
still enjoy extraordinary popularity due to the wealth of their physical applications and
connections to mathematics,  which are described in the surveys \cite{Nekr, Per,Banff,Suth}.
In correspondence to the many variants of these systems (associated with different
interaction potentials,
root systems, relativistic deformations, inclusion of two types
of particles,  and so on) there exist also
several approaches to studying them.
The systems based on trigonometric/hyperbolic interaction potentials
are usually called Sutherland type and here we study a particular case of such systems
in the classical Hamiltonian reduction framework, reviewed for example in \cite{Per}.

A Sutherland type integrable system  describing the interaction
of $m$ `positively charged' and
$(n-m)$ `negatively charged' particles was first introduced by Calogero
\cite{Cal} by means
of shifting the positions of $m$ out of the $n$ particles by  $\ri \frac{\pi}{2}$.
This trick converts the repulsive interaction potential $\sinh^{-2}(q_j-q_k)$ into the attractive
potential $-\cosh^{-2}(q_j-q_k)$ between the particles of opposite charge (indexed say by
$1\leq j \leq m < k \leq n$).
Then Olshanetsky and Rogov \cite{OR} derived
the Calogero-Sutherland model by Hamiltonian reduction of free
motion on an affine symmetric space.
The dynamics of the model and its relativistic deformation  was
analyzed in detail by Ruijsenaars \cite{RIMS94}, motivated mainly by the relation of this problem
to the interaction of sine-Gordon  solitons and anti-solitons.

In a little noticed paper Hashizume \cite{Ha}  generalized the Olshanetsky-Rogov derivation and
thereby proved the integrability of a family of hyperbolic Sutherland models associated
to the so-called root systems with signature.
In the case of the $BC(n)$ root system, his model involves two types of particles
moving on the half-line with the interaction governed by two independent coupling parameters.

It is well-known that integrable $BC(n)$ Sutherland models involve in general three arbitrary
couplings corresponding to the three different root lengths.
However, it has been explained only rather recently, by Pusztai and one of us \cite{FP1},
how the three couplings  arise in the setting of Hamiltonian reduction.
In the present paper, we generalize the result of \cite{FP1} and derive the following
$BC(n)$ type Sutherland Hamiltonian,
\bea
&& H= \frac{1}{2} \sum_{j=1}^n p_j^2
- \sum_{1\leq j\leq  m<k \leq n} (\frac{\kappa^2}{\cosh^2(q_j - q_k)} +
\frac{\kappa^2}{\cosh^2(q_j + q_k)} )
\nonumber\\
&&\phantom{X}
+ \sum_{1\leq j < k \leq m} ( \frac{\kappa^2 }{\sinh^2(q_j - q_k)} +
\frac{\kappa^2 }{\sinh^2(q_j + q_k)} )
+ \sum_{m<j< k\leq n} (\frac{\kappa^2}{\sinh^2(q_j - q_k)} + \frac{\kappa^2}{\sinh^2(q_j + q_k)} )
\nonumber\\
&&\phantom{XXX}
+ \frac{1}{2}\sum_{j=1}^n \frac{(x - y)^2}{\sinh^2(2 q_j)}
+ \frac{1}{2}\sum_{j=1}^m \frac{xy}{\sinh^2(q_j)} -
\frac{1}{2}\sum_{j=m+1}^n \frac{  xy }{\cosh^2( q_j)},
\label{I1}\eea
where $\kappa>0$, $x$ and $y$ are real coupling constants.
If $(x^2 - y^2)\neq 0$, then energy conservation
ensures that the corresponding dynamics can be consistently restricted
to the domain where
\be
q_1>q_2>...>q_m>0\quad\hbox{and}\quad
\quad
q_{m+1}> q_{m+2} >...> q_n>0.
\ee
Supposing also that $xy>0$,
the Hamiltonian (\ref{I1}) describes attractive-repulsive
 interactions between $m$ `positively charged' and $(n-m)$ `negatively charged'
particles influenced also by their mirror images and a positive charge
fixed at the origin.

We shall derive
the model (\ref{I1}) by reduction of the free geodesic motion
on the group $G= SU(n,n)$.
Our reduction relies on a symmetry group of the form
$G_+ \times G^+$, where $G_+<G$ is a maximal compact subgroup and $G^+<G$ is
the (non-compact)
 fixed point subgroup of a $G$-involution that commutes with the
Cartan involution fixing $G_+$.
Hashizume \cite{Ha} reduced the geodesic motion on affine symmetric spaces
such as $G/G^+$, which itself is the reduction of the free  motion on $G$
associated with the zero value of
the moment map of the $G^+$  symmetry.
He obtained the model (\ref{I1}) with two independent
couplings, while we obtain  it with three
 arbitrary couplings since we use non-trivial
one-point coadjoint orbits  of $G^+$ having a free parameter
(corresponding to $y$ in (\ref{I1})) to define our reduction.
In \cite{FP1} (see also \cite{Obl,FP2}) the symmetry
 group $G_+ \times G_+$ was used
in an analogous manner to describe the $m=0$ case.

To be more precise regarding the comparison with Ref.~\cite{Ha}, note that
the model (\ref{I1}) with $y=0$ was obtained in \cite{Ha}
by using $G=U(n,n)$ and the model with a certain non-linear relation between
the three couplings was obtained by using $G=U(n+1,n)$.
The possible alternative reduction treatments
of the model (\ref{I1}) are briefly discussed also in the concluding section.

Our derivation implies the Liouville integrability of the model (\ref{I1})
in the general case, and it also gives rise to a simple linear-algebraic
algorithm for constructing the solutions.
It could be interesting to analyze the dynamics of the model
in the future by utilizing this algorithm, and to possibly relate
it to special solutions
in a field theory on the half-line.
Further comments on open problems are offered at the end of the paper.

\section{Group theoretic preliminaries}
\setcounter{equation}{0}

We here fix our notations and recall some group theoretic results
that will be needed later.

To begin, we choose some integers
\be
1\leq m < n
\label{2.}\ee
and define the matrices
\be
Q_{n,n}:=\left[
\begin{array}{cc}
0 & \1_n\\
\1_n & 0
\end{array}\right]\in gl(2n,\bC),
\quad
I_{m}:=\operatorname{diag}(\1_m,-\1_{n-m})
\in gl(n,\mathbb{C}),
\label{2.2}\ee
where $\1_n$ denotes the $n\times n$ unit matrix. We also introduce
\be
D_m := \operatorname{diag}(I_{m}, I_{m}) =
\operatorname{diag}(\1_m, -\1_{n-m}, \1_{m}, -\1_{n-m})
\in gl(2n,\bC).
\label{2.3}\ee
We adopt the convention in which the group $G:= SU(n,n)$ and its Lie algebra $\cG:= su(n,n)$
are given by
\be
SU(n,n) = \{ g \in SL(2n,\bC)\,\vert\, g^\dagger Q_{n,n} g = Q_{n,n}\}
\label{2.4}\ee
and
\be
su(n,n) = \{ V \in sl(2n,\bC)\,\vert\, V^\dagger Q_{n,n} + Q_{n,n} V =0\}.
\label{2.5}\ee
In the obvious $n\times n$ block notation the elements $V\in su(n,n)$ have the form
\be
V=
\left[
\begin{array}{cc}
X & Y\\
Z & -X^\dagger
\end{array}\right],
\quad
Y^\dagger = -Y, \quad Z^\dagger = -Z,
\quad
\Im(\oper{tr}(X))=0.
\label{2.6}\ee

We consider the commuting involutions of $SU(n,n)$ provided by the Cartan involution $\Theta$
and the involution $\Gamma$:
\be
\Theta(g):= (g^\dagger)^{-1},
\qquad
\Gamma(g):= D_m \Theta(g) D_m,
\qquad
\forall g\in G.
\label{2.7}\ee
The fixed point subgroup  of $\Theta$ is the maximal compact subgroup $G_+ <G$
 and the (non-compact) fixed point subgroup
of $\Gamma$ is
denoted by $G^+$.
Let $\theta$ and $\gamma$ be the corresponding involutions of $\cG = su(n,n)$.
Using the $n\times n$ block notation,
the Lie algebra $\cG_+$ of $G_+$ reads
\be
\cG_+=\left\{
\left[
\begin{array}{cc}
X & Y\\
Y & X
\end{array}\right]:
\,\,
X^\dagger = -X, \,\, Y^\dagger = -Y, \,\, \oper{tr}(X)=0\right\},
\label{2.8}\ee
and is isomorphic to $s(u(n) \oplus u(n))$ according to
\be
s(u(n) \oplus u(n))\ni \left[
\begin{array}{cc}
\alpha & 0\\
0 & \beta
\end{array}\right] \mapsto \psi(\alpha,\beta):=
\frac{1}{2}\left[
\begin{array}{cc}
\alpha+\beta & \alpha-\beta\\
\alpha-\beta & \alpha+\beta
\end{array}\right]\in \cG_+.
\label{2.9}\ee
Correspondingly, the group $G_+$ is isomorphic to $S(U(n) \times U(n))$
 via the formula
\be
S(U(n) \times U(n))\ni \left[
\begin{array}{cc}
a & 0\\
0 & b
\end{array}\right] \mapsto
g(a,b):=
  \frac{1}{2}\left[
\begin{array}{cc}
a+b & a-b\\
a-b & a+b
\end{array}\right]\in G_+,
\label{2.9+}\ee
which can be written also as
\be
g(a,b) = K (\oper{diag}(a,b)) K^{-1}
\quad
\hbox{with}\quad
K:= \frac{1}{\sqrt{2}} \left[
\begin{array}{cc}
\1_n & \1_n\\
\1_n & -\1_n
\end{array}\right].
\label{C}\ee

The Lie algebra $\cG^+$ of $G^+$ is furnished by
\be
\cG^+= \left\{
\left[
\begin{array}{cc}
X & Y\\
I_{m}Y I_{m}  & I_{m}X I_{m}
\end{array}\right]:
\,\,
X^\dagger = -I_{m}X I_{m}, \,\, Y^\dagger = -Y, \,\, \oper{tr}(X)=0\right\},
\label{2.10}\ee
and is isomorphic  to $s(u(m,n-m) \oplus u(m,n-m))$ via the map
\be
s(u(m,n-m) \oplus u(m,n-m))\ni \left[
\begin{array}{cc}
\alpha & 0\\
0 & \beta
\end{array}\right] \mapsto \chi(\alpha,\beta):=
\frac{1}{2}\left[
\begin{array}{cc}
\alpha+\beta & (\alpha-\beta)I_m\\
I_m(\alpha-\beta) & I_m(\alpha+\beta)I_m
\end{array}\right]\in \cG^+.
\label{E1}\ee
In the above formula $u(m,n-m)$ is realized as the Lie algebra of
the $n \times n$ matrices satisfying the relation
\be
\alpha^\dagger I_m + I_m \alpha = 0
\label{E2}\ee
and it holds that
\be
\chi(\alpha,\beta)
 = {\tilde K} (\oper{diag}(\alpha,\beta)) {\tilde K}^{-1}
\quad\hbox{with}\quad \tilde K:= \frac{1}{\sqrt{2}} \left[
\begin{array}{cc}
\1_n & \1_n\\
I_m & - I_m
\end{array}\right].
\ee
Similarly,  $G^+$ is isomorphic to
$S( U(m,n-m) \times U(m,n-m))$ by means of conjugation by $\tilde K$.

The eigensubspaces $\cG_-$ of $\theta$ and $\cG^-$  of $\gamma$
having eigenvalue $-1$ can be displayed  as
\be
\cG_- =\left\{
\left[
\begin{array}{cc}
X & Y\\
-Y & -X
\end{array}\right]:
\,\,
X^\dagger = X, \,\, Y^\dagger = -Y\right\}
\label{2.11}\ee
and
\be
\cG^-=
\left\{\left[
\begin{array}{cc}
X & Y\\
-I_mYI_m & -I_m X I_m
\end{array}\right]:
\,\,
X^\dagger = I_m X I_m, \,\, Y^\dagger = -Y\right\}.
\label{2.12}\ee
Introducing $\cG_s^r:= \cG_s \cap \cG^r$ for any signs $s,r \in \{\pm\}$, we can
decompose $\cG$ as the direct sum of disjoint subspaces,
\be
\cG = \cG_-^- \oplus \cG_-^+ \oplus \cG_+^- \oplus \cG_+^+,
\label{2.13}\ee
which are pairwise perpendicular to each other
 with respect to the invariant scalar product on $\cG$ defined by
\be
\langle V, W \rangle := \frac{1}{2} \tr (V W),
\qquad
\forall V, W \in \cG.
\label{2.14}\ee

We have
\be
\cG_+^+=\left\{
\left[
\begin{array}{cc}
X & Y\\
Y & X
\end{array}\right]:
\,\,
X = I_m X I_m=-X^\dagger, \,\, Y=I_mY I_m=-Y^\dagger, \,\, \oper{tr}(X)=0\right\},
\label{G++}\ee
and the appropriate  restriction of the map (\ref{2.9}) gives rise to an isomorphism
\be
\cG_+^+\simeq s((u(m)\oplus u(n-m)) \oplus (u(m)\oplus u(n-m)) ).
\ee
We shall also use the explicit form of the other subspaces:
\bea
&&\cG_+^-=\left\{
\left[
\begin{array}{cc}
X & Y\\
Y & X
\end{array}\right]:
\,\,
X = -I_m X I_m=-X^\dagger, \,\, Y=-I_mY I_m=-Y^\dagger \right\},
\nonumber\\
&&\cG_-^+=\left\{
\left[
\begin{array}{cc}
X & Y\\
-Y & -X
\end{array}\right]:
\,\,
X = -I_m X I_m=X^\dagger, \,\, Y=-I_mY I_m=-Y^\dagger \right\},\nonumber\\
&&\cG_-^-=\left\{
\left[
\begin{array}{cc}
X & Y\\
-Y & -X
\end{array}\right]:
\,\,
X = I_m X I_m=X^\dagger, \,\, Y=I_mY I_m=-Y^\dagger \right\}.
\label{2.15+}\eea

Next, for our purpose we choose a maximal Abelian subspace $\cA$ of $\cG^-_-$,
i.e.,  an Abelian subalgebra of $\cG$ which lies in $\cG_-^-$ and
is not properly contained in any Abelian subalgebra of the same kind.
It is known \cite{KAH} that
any two choices are equivalent by the conjugation action of $G_+^+ $ on $\cG_-^-$,
and concretely we choose
\be
\cA:=
\left\{ q:=\left[
\begin{array}{cc}
\mathbf{q} & 0\\
0 & -\mathbf{q}
\end{array}\right]:
\,\,
\mathbf{q}=\oper{diag}(q_1,..., q_n),\,\, q_k\in \bR\right\}.
\label{2.15}\ee
One can verify that
the centralizer of $\cA$ in $\cG$ is given by the direct sum
\be
\cC = \cA \oplus \cM,
\quad
\cM=\left\{ d:=\ri \left[
\begin{array}{cc}
\mathbf{d} & 0\\
0 & \mathbf{d}
\end{array}\right]:
\,\,
\mathbf{d}=\oper{diag}(d_1,..., d_n),\,\, d_k\in \bR,\,\, \tr(d) =0\right\}< \cG_+^+.
\label{2.16}\ee
Denote by $A$ and $M$ the connected subgroups of $G$ corresponding to the Abelian
subalgebras $\cA$ and $\cM$, respectively.
In fact, $M$ is precisely the subgroup of $G_+^+$ whose elements $g$ satisfy
\be
g q g^{-1} = q,
\qquad
\forall q\in \cA.
\label{2.17}\ee
Furthermore, we call an element $q\in \cA$ \emph{regular} if the elements $g\in G_+^+$ satisfying
the relation $g q g^{-1}=q$ all belong to $M$.
It is not difficult to check that $q\in \cA$ is regular in this sense if and only
if the following conditions hold:
\be
q_i \neq 0 \quad i=1,..., n,
\quad (q_j - q_k) (q_j + q_k) \neq 0
\quad  1\leq j < k\leq m
\quad\hbox{and}
\quad  m<j < k\leq n.
\label{qreg}\ee

Choose a connected component  $\cA_c$ of the open set formed by regular elements
of $\cA$, and denote the closure of
this `open Weyl chamber' by $\bar\cA_c$.
According to general results \cite{KAH}, every element $g\in G$ can be decomposed in the form
\be
g = g_+ e^q g^+
\quad\hbox{with}\quad q\in \bar\cA_c,\, g_+\in G_+,\, g^+ \in G^+.
\label{2.18}\ee
The constituent $q$ that enters this decomposition is unique, and if $q$ is regular then the
ambiguity  of the pair $(g_+, g^+)$ is exhausted by the replacement
\be
(g_+, g^+) \to (g_+ \mu, \mu^{-1} g^+)
\qquad
\forall \mu \in M.
\label{2.19}\ee
In the the generalized Cartan decomposition (\ref{2.18}) the
open Weyl chamber $\cA_c$ can be taken to consist of the elements $q$
in (\ref{2.15}) that are subject to the condition
\be
q_1 > q_2 > ... > q_m >0
\quad\hbox{and}\quad
q_{m+1} > q_{m+2} > ...> q_n >0.
\label{2.20}\ee

Both $\cG_+$ and $\cG^+$ possess one-dimensional centres.
The centre of $\cG_+$ is generated  by
\be
C^l:= \ri  Q_{n,n} =\ri \left[\begin{array}{cc}
0 & \1_n\\
 \1_n & 0
\end{array}\right]
\label{2.21}\ee
and the centre of $\cG^+$ is spanned  by
\be
C^r:= \ri  \left[\begin{array}{cc}
0 & I_m\\
 I_m & 0
\end{array}\right].
\label{2.22}\ee
These elements enjoy the property
\be
C^\lambda \in \cM^\perp \cap \cG^+_+
\quad\hbox{for}\quad \lambda =l, r.
\label{2.23}\ee
The decomposition (\ref{2.18}) and the property
(\ref{2.23}) will be important for us in Section 3.

By means of the invariant scalar product (\ref{2.14}),
we can regard $\cG$, $\cG_+$ and $\cG^+$ as their own dual spaces, respectively.
This then also identifies the respective coadjoint actions with the
adjoint actions.
In the next section we shall utilize  particular coadjoint orbits of $G_+$.
To describe them,
for any non-zero column vector $u\in \bC^n$ define
the matrices
\be
X(u) := \ri (u u^\dagger - \frac{u^\dagger u}{n} \1_n)
\quad\hbox{and}\quad
\xi(u):=
\frac{1}{2}
\left[
\begin{array}{cc}
X(u) & X(u)\\
X(u) & X(u)
\end{array}\right].
\label{2.24}\ee
Fixing arbitrary real constants $\kappa>0$ and $x\neq 0$,
it is easy to see (cf.~(\ref{2.9})-(\ref{C})) that the set
\be
\cO_{\kappa,x} :=
\{x C^l +\xi(u) \vert\, u\in \bC^n, \,  u^\dagger u = 2 \kappa n \}
\label{2.25}\ee
is
a coadjoint orbit of $G_+$ of minimal non-zero dimension.
The action of $g(a,b) \in G_+$ on $\cO_{\kappa,x}$ takes the form
\be
g(a,b) (x C^l +\xi(u)) g(a,b)^{-1} = (xC^l + \xi(au)).
\ee
Since $\xi(u)$ determines $u$ up to an overall $U(1)$ phase,
the orbit $\cO_{\kappa,x}$ can be identified with the
the complex projective space $\bC P_{n-1}$.

We remark that, for any real constants $x$ and $y$,  $x C^l $ and $y C^r$
represent one-point coadjoint orbits of $G_+$ and $G^+$, respectively.

\section{Hamiltonian reduction}
\setcounter{equation}{0}

We shall reduce the free geodesic motion
on the group $G=SU(n,n)$ formulated as a Hamiltonian system on
the cotangent bundle $T^* G$.
We find it convenient to analyze the reduction
by using the so-called shifting trick of symplectic geometry, which amounts to extending the
phase space by a coadjoint orbit before reduction \cite{OrtRat}.
Specifically,   trivializing $T^* G$ by right-translations and identifying
$\cG^*$ with $\cG$ by means of the invariant scalar product,
we consider the phase space
\be
P:= T^* G \times \cO_{\kappa, x} \simeq (G\times \cG)\times \cO_{\kappa, x} \equiv
\{ (g, J, \zeta)\, \vert\, g\in G, \, J\in \cG,\, \zeta\in
\cO_{\kappa, x} \}.
\label{3.1}\ee
The symplectic form on $P$ is given by
\be
\Omega = \Omega_{T^*G} + \Omega_{\cO_{\kappa, x}},
\label{Om}\ee
where $\Omega_{T^*G}$ can be written explicitly as
\be
\Omega_{T^*G}  = d \langle J, dg g^{-1} \rangle
\label{3.2}\ee
while the explicit form of
the Kirillov-Kostant-Souriau symplectic form $\Omega_{\cO_{\kappa, x}}$ of the coadjoint
orbit $\cO_{\kappa, x}$ (\ref{2.25}) will not be needed.
The phase space $P$ carries the commuting family of Hamiltonians
provided by
\be
H_k(g,J,\zeta):= \frac{1}{4k}  \tr(J^{2k}),
\qquad
k=1,2,..., n,
\label{Hk}\ee
the first member of which is responsible for the geodesic motion.
These Hamiltonians are explicitly integrable;
the flow of $H_k$ with initial value $(g_0, J_0, \zeta_0)$ is readily verified to
be
\be
(g(t), J(t), \zeta(t)) = (e^{t V_{k}} g_0, J_0, \zeta_0)
\quad\hbox{with}\quad
V_{k}:= J_0^{2k-1} - \frac{1}{2n} \tr (J_0^{2k-1}) \1_{2n}.
\label{freesol}\ee
Note that $H_k$ is real since $J^{2k}$ satisfies
$(J^{2k})^\dagger = Q_{n,n} J^{2k} (Q_{n,n})^{-1}$, and $V_{k}$ in (\ref{freesol}) belongs to
$\cG=su(n,n)$.

We introduce an action of the group $G_+ \times G^+$ on $P$ by sending
the pair $(\eta, h)\in G_+ \times G^+$ to the symplectomorphism $\Psi_{\eta, h}$ of $P$
operating as follows:
\be
\Psi_{\eta, h}(g,J,\zeta):= (\eta g h^{-1}, \eta J \eta^{-1}, \eta\zeta \eta^{-1}).
\label{3.5}\ee
The Hamiltonians $H_{k}$ (\ref{Hk}) are invariant under this
 group action, which is generated by the equivariant moment map
\be
\Phi = (\Phi_+, \Phi^+): P \to (\cG_+, \cG^+),
\label{3.6}\ee
\be
\Phi_+(g, J, \zeta) = \pi_+(J) + \zeta,
\qquad
\Phi^+(g, J, \zeta)= - \pi^+(g^{-1} J g),
\label{3.7}\ee
where the projection $\pi_+:\cG \to \cG_+$ is
given by means of the decomposition $\cG = \cG_+ \oplus \cG_-$ and
 $\pi^+: \cG \to \cG^+$   by $\cG= \cG^+ \oplus \cG^-$.

We are interested in the reduction defined by imposing the moment map constraint
\be
\Phi=\nu
\quad
\hbox{with}\quad
\nu := (0, -y C^r),
\label{3.8}\ee
where $y\neq 0$ is a real constant and we refer to (\ref{3.6}).
The action of the symmetry group $G_+ \times G^+$ preserves
 the `constraint surface'
\be
 P_c:= \Phi^{-1}(\nu)\subset P.
\label{Pc} \ee
We require that the constants $x$ and $y$ verify
\be
(x^2 - y^2) \neq 0.
\label{xy}\ee
Then the corresponding space of orbits,
\be
P_\red:= P_c / (G_+ \times G^+),
\label{3.10}\ee
will turn out to be a smooth manifold.
According to the general theory \cite{OrtRat},
$P_\red$ inherits the symplectic form $\Omega_\red$ and the
reduced Hamiltonians $H_{k}^\red$
defined by the formulas
\be
\pi^* \Omega_\red = \Omega\vert_{P_c},
\qquad
\pi^* H_{k}^\red = H_{k}\vert_{P_c},
\label{3.11}\ee
where $\pi: P_c \to P_\red$ is the natural projection and $\Omega\vert_{P_c}$ is
the restriction of $\Omega$ (\ref{Om}) on $P_c \subset P$.

\medskip
\noindent
{\bf Remark 3.1.}
In this technical remark we explain why the space of orbits (\ref{3.10}) is a smooth manifold.
First, we note that the action (\ref{3.5}) of
$G_+ \times G^+$ is  \emph{proper}.
By definition \cite{OrtRat},
this means  that for any sequences
$(\eta_n, h_n)$ in  $G_+ \times G^+$  and
$(g_n,J_n,\zeta_n)$ in  $P$ (with $n\in \bN)$ for which
$(g_n,J_n,\zeta_n)$ and $\Psi_{(\eta_n, h_n)}(g_n,J_n,\zeta_n)$ are both convergent, there
exists a convergent subsequence of the sequence $(\eta_n, h_n)$.
To show this, choose a convergent subsequence  $\eta_{n_i}$ of the sequence
$\eta_n$ in $G_+$.
This is always possible since $G_+$ is compact.
Then, by considering the convergent sequences $\eta_{n_i} g_{n_i} (h_{n_i})^{-1}$ and
$g_{n_i}$ one can immediately conclude that $h_{n_i}$ must be a convergent sequence in $G^+$, which
proves the claim.
To continue, notice from (\ref{3.5})  that the effectively acting symmetry group is
the factor group
 $(G_+ \times G^+)/ (\bZ_n)_\diag$, where
 $(\bZ_n)_\diag$ is formed by the pairs $ (z \1_{2n}, z \1_{2n}) \in G_+ \times G^+$ with $z$ running
 over the $n^{\mathrm{th}}$ roots of unity.
 We shall demonstrate in the proof of Theorem 3.4  that
 the action of $(G_+ \times G^+)/ (\bZ_n)_\diag$ on $P_c$ is a \emph{free} action.
Moreover, $P_c$ is
 a closed, embedded submanifold of $P$, as it follows from the definition (\ref{Pc}) of $P_c$
 and from the locally free character of the $(G_+\times G^+)$-action on it.
 Since we have a free and proper action on the manifold $P_c$,  the general theory \cite{OrtRat}
 guarantees that $P_\red\simeq P_c/ ((G_+ \times G^+)/ (\bZ_n)_\diag)$
 is  a smooth symplectic manifold.
 This is manifest
by the model of $P_\red$ constructed below.

\medskip

Our goal in what follows is to exhibit a global cross section (a global `gauge slice')
of the orbits of $G_+ \times G^+$ in $P_c$, which will yield a concrete model
of the reduced Hamiltonian systems $(P_\red, \Omega_\red, H_{k}^\red)$.
We first present the following lemma, whose proof will also show
how to construct a convenient global gauge slice.

\medskip
\noindent
{\bf Lemma 3.2.}
\emph{The element $e^q$, $q\in \cA$ in (\ref{2.15}), and $u\in \bC^n$
enter a triple
$(e^q,J, x C^l + \xi(u))\in P_c$ (\ref{Pc}) if and only if
$\vert u_j\vert^2 = 2 \kappa$ for all $j=1,..., n$
and $q$ is regular in the sense of
Eq.~(\ref{qreg}).}
\begin{proof}
Let us inspect the moment map constraint for an element of $P$ of the form
\be
(e^q, J, x C^l + \xi(u))
\quad\hbox{with some}\quad
q\in \cA.
\label{3.12}\ee
Denoting the projections associated to the decomposition (\ref{2.13}) as $\pi^r_s$
and decomposing $J$ as
\be
J= J_+^+ + J_+^- + J^+_- + J_-^-,
\label{3.13}\ee
we can spell out the moment map constraint as the conditions
\be
J_+^+ = - x C^l - \pi_+^+ (\xi(u)),
\qquad
J_+^- = - \pi_+^-(\xi(u)),
\label{3.14}\ee
and
\be
\pi^+(e^{-\ad_q}(J))\equiv(\cosh\ad_q)(J_+^+ + J_-^+) - (\sinh \ad_q)(J_+^- + J_-^-) = y C^r.
\label{3.15}\ee
Since $C^r \in \cG_+^+$, the $\pi^+_-$ projection of  equation (\ref{3.15}) says that
\be
(\cosh\ad_q)( J_-^+) - (\sinh \ad_q)(J_+^-) = 0,
\label{3.16}\ee
and its $\pi^+_+$ projection requires that
\be
(\cosh\ad_q)(J_+^+ ) - (\sinh \ad_q)(J_-^-) = y C^r.
\label{3.17}\ee
We here used that $\cosh\ad_q$ maps $\cG_s^r$ to $\cG_s^r$ and
$\sinh\ad_q$ maps $\cG_s^r$ to $\cG_{-s}^{-r}$ (with $-s= \mp$ for $s=\pm$).
By substituting $J_+^+$ from (\ref{3.14}) into (\ref{3.17}) and then
 taking the scalar product of both sides of equation (\ref{3.17}) with an
 arbitrary $T\in \cM$ (\ref{2.16}),
  we obtain  the requirement
 \be
 \langle T, \xi(u) \rangle =0
 \qquad
 \forall T\in \cM,
 \label{3.18}\ee
 where we also took into account that $C^l$ and $C^r$ belong to
 $\cM^\perp$ (\ref{2.23}).
By using the form of $\cM$ (\ref{2.16}) and that of $\xi(u)$ (\ref{2.24}),
we can rewrite (\ref{3.18}) as
the condition
\be
\vert u_j \vert^2 = 2\kappa,
\qquad
\forall j=1,..., n.
\label{3.19}\ee

If (\ref{3.19}) holds, then we can apply the action of the subgroup
$M_\diag$ of $G_+ \times G^+$,
\be
M_\diag:= \{(\mu, \mu) \in G_+ \times G^+ \,\vert\, \mu \in M\}
\label{3.20}\ee
to replace (without changing $q$) the element in (\ref{3.12})
by an element of the form
\be
(e^q, J, xC^l+ \xi(u^\kappa))
\quad\hbox{with the vector}\quad
u^\kappa_j := \sqrt{2\kappa},
\qquad
j=1,..., n.
\label{3.21}\ee
We further inspect the moment map constraint for the element (\ref{3.21}).
First looking at
the block-diagonal components of (\ref{3.17}), we see that
the matrix elements
 ${(J_-^-)}_{k,k}$
 are arbitrary real numbers
for all $k=1,..., n$, and that we must have
\be
-\ri \kappa \cosh (q_j - q_k) - {(J_-^-)}_{j,k} \sinh(q_j - q_k) =0
\quad
\hbox{for}\quad
1\leq j < k \leq m
\quad\hbox{and}\quad
m<j<k \leq n.
\label{3.22}\ee
The last equation can be solved for ${(J_-^-)}_{j,k}$ if and only if $(q_j - q_k)\neq 0$ for the
pertinent indices.
Next,
the block off-diagonal components of (\ref{3.17}) can be spelled out as the conditions
\be
-\ri \kappa  \cosh (q_j + q_k) - {(J_-^-)}_{j,n+k} \sinh(q_j + q_k) =0
\quad
\hbox{for}\quad
1\leq j < k \leq m
\quad\hbox{and}\quad
m<j<k \leq n,
\label{3.23}\ee
and
\be
- \ri x \cosh(2q_k)  - {(J_-^-)}_{k,n+k} \sinh(2q_k) = y(C^r)_{k,n+k}
\quad \hbox{for}\quad k=1,..., n.
\label{3.24}\ee
Equation (\ref{3.23}) can be solved for ${(J_-^-)}_{j,n+k}$ if and only if
$(q_j + q_k) \neq 0$ for the relevant indices.
Taking into account the assumption $(x^2 - y^2)\neq 0$ (\ref{xy}) and the formula
(\ref{2.22}) of $C^r$,
equation (\ref{3.24}) can be solved for ${(J_-^-)}_{k,n+k}$ if and only if $q_k\neq 0$ for all $k$.

We have seen that equation  (\ref{3.17}) admits a solution if and only if $u$ satisfies
(\ref{3.19}) and $q$ is regular (\ref{qreg}). The proof is finished by noting
that the remaining equation (\ref{3.16}) can always be solved
for $J_-^+$ if $J_+^- =- \pi_+^-(\xi(u))$ is given, since
$\cosh \ad_q$ yields an  invertible map on $\cG_-^+$.
\end{proof}

\medskip
\noindent
{\bf Definition 3.3.}
\emph{
Suppose that $\kappa>0$ and $x,y$ satisfy (\ref{xy}).
For any $q\in \cA_c$ and $p\in \cA$ define the function $J(q,p)$ by the formula
\be
J(q,p) := - xC^l - \xi(u^\kappa) + L(q,p),
\label{Jqp}\ee
where $(u^\kappa)_j=\sqrt{2\kappa}$ ($j=1,..., n$) and the matrix elements of $L(q,p)= \pi_-(J(q,p))$ are
the following.
Firstly, if $1\leq j < k \leq m$ or $m<j<k \leq n$, then
\be
L_{j,k} = -L_{k,j}=-L_{j+n, k+n}=L_{k+n,j+n}  = -\ri \kappa \coth(q_j-q_k),
\ee
\be
L_{j,k+n} = L_{k,j+n}=-L_{j+n, k}=-L_{k+n,j}  = -\ri \kappa \coth(q_j+q_k).
\ee
Secondly, if $1\leq j \leq m$ and $m<k \leq n$, then
\be
L_{j,k} = -L_{k,j}=-L_{j+n, k+n}=L_{k+n,j+n}  = -\ri \kappa \tanh(q_j-q_k),
\ee
\be
L_{j,k+n} = L_{k,j+n}=-L_{j+n, k}=-L_{k+n,j}  = -\ri \kappa \tanh(q_j+q_k).
\ee
Finally, for any $1\leq j\leq m$,  $m<k\leq n$, and $1\leq l \leq n$, we have
\be
L_{j,j+n} = - L_{j+n, j}=-\frac{\ri y}{\sinh (2q_j)} - \ri x \coth(2q_j),
\ee
\be
L_{k,k+n}=- L_{k+n,k} = \frac{\ri y}{\sinh (2q_k)} - \ri  x \coth(2q_k).
\ee
\be
L_{l,l} =- L_{l+n,l+n}= p_l.
\ee
}

\medskip
\noindent
{\bf Theorem 3.4.} \emph{By using the above definition of $J(q,p)$, consider the set
\be
S= \{ (e^q, J(q,p), xC^l + \xi(u^\kappa))\,\vert\, q\in \cA_c,\,p\in \cA\,\}.
\label{S}\ee
The submanifold $S\subset P$ lies in the constraint surface $P_c$ (\ref{Pc}) and intersects
every orbit of $G_+ \times G^+$ in $P_c$ precisely in one point.
The pull-back  $\Omega_S$ of the symplectic form $\Omega$ (\ref{Om}) on $S$ is given by
\be
\Omega_S =\sum_{k=1}^n d p_k \wedge d q_k.
\label{OmS}\ee
Thus the symplectic manifold $(S, \Omega_S)$ provides a model
of the reduced phase space $(P_\red, \Omega_\red)$ (\ref{3.10}), which can be identified
with the cotangent bundle
$T^* \cA_c$.}

\begin{proof}
We know that every $g\in G$ can be decomposed according to (\ref{2.18}),
and Lemma 3.2 implies that every gauge orbit (i.e.~$G_+\times G^+$ orbit) in $P_c$
admits a representative of the form
\be
(e^q, J, xC^l + \xi(u^\kappa))
\quad
\hbox{with}\quad
q\in \cA_c,
\label{3.37}\ee
where $\cA_c$ is an open Weyl chamber (for example the one defined in (\ref{2.20})).
Following the proof of Lemma 3.2, it is easy to check  that $J$ in (\ref{3.37})
can be written as $J= J(q,p)$ in (\ref{Jqp}) with some $p\in \cA$. Indeed,  the formula
(\ref{Jqp}) was obtained by directly solving
the constraint equations listed in the proof of Lemma 3.2.
To check that $S$ intersects every gauge orbit only once, suppose that we have
\be
(\eta e^q h^{-1},\eta J(q,p) \eta^{-1}, x C^l + \eta \xi(u^\kappa) \eta^{-1}) =
(e^{q'}, J(q', p'),xC^l+ \xi(u^\kappa)),
\quad
(\eta, h) \in G_+ \times G^+,
\label{3.38}\ee
for two triples in $S$.
The uniqueness property of the decomposition (\ref{2.18})
entails that
$e^q= e^{q'}$, which is equivalent to $q=q'$, and $(\eta,h) = (\mu,\mu)$ for some
$\mu \in M$. Then it follows from the second component of the equality in (\ref{3.38}) that $p=p'$  holds,
i.e., the two representatives of the orbit coincide.
Incidentally, the equality $\mu \xi(u^\kappa) \mu^{-1} = \xi(u^\kappa)$
implies that  $\mu\in M$ must belong to the centre of  $M$, which
is isomorphic to the group $\bZ_n$ and equals the centre of $G$.
The corresponding subgroup $(\bZ_n)_\diag < M_\diag$ acts trivially on $P$, and hence
we can also conclude that the factor group $(G_+ \times G^+)/ (\bZ_n)_\diag$ acts
freely on the constraint surface $P_c$.

We can  compute the pull-back of $\Omega$ (\ref{Om}) on the submanifold
$S\subset P$, which gives the formula (\ref{OmS}). Since we have seen that
$S$ is a global cross section of the gauge orbits in $P_c$, it follows that
$(S,\Omega_S)$ represents a model of the reduced phase space $(P_\red,\Omega_\red)$.
Referring to the identification $\cA \simeq \cA^*$ defined by the scalar product
(\ref{2.14}) of $\cG$,
 $(S,\Omega_S)$ is  symplectomorphic to the
cotangent bundle $T^*\cA_c \simeq \cA_c \times \cA^*$ equipped with the Darboux symplectic form.

\end{proof}

Let us recall that
a Hamiltonian given by a smooth function on a $2n$ dimensional symplectic manifold
is called
\emph{Liouville integrable}
if it is contained in a family of $n$ functionally independent, globally smooth functions
on the phase space
whose mutual Poisson brackets vanish and their Hamiltonian flows
are complete.
Now the following result is an immediate consequence of the Hamiltonian reduction.

\medskip
\noindent
{\bf Corollary 3.5.}
\emph{A family of functionally independent Hamiltonians that are in involution with respect to the
canonical Darboux Poisson structure on $T^* \cA_c$ is provided by
\be
H_{k}^\red = \frac{1}{4k} \tr (J(q,p)^{2k}),
\quad
k=1,..., n.
\label{3.39}\ee
The  generalized Sutherland
Hamiltonian $H(q,p)$ (\ref{I1}) is Liouville integrable, since it obeys
\be
H(q,p) =\frac{1}{4} \tr( J(q,p)^2)=H_1^\red(q,p).
\label{3.40}\ee}
\medskip

\begin{proof}
The reduced Hamiltonians (\ref{3.39}) are in involution with respect to the canonical
Poisson structure
derived from $\Omega_S$ (\ref{OmS}) since the original Hamiltonians $H_{k}$ (\ref{Hk}) are in involution
with respect to the Poisson structure on $(P, \Omega)$.
By using Definition 3.3, the identity (\ref{3.40}) is a matter of direct verification.
At generic points of the phase space, the Hamiltonians (\ref{3.39}) are independent,
since they start with independent
`leading terms' given by respective homogeneous polynomials in $p_1,..., p_n$.
The reduction guarantees that the corresponding Hamiltonian flows are complete, and
 thus  $H_{k}^\red$ (and in particular $H=H_1^\red$) is Liouville integrable.
\end{proof}

Finally, let us describe how the flows of the reduced Hamiltonians $H_k^\red$ can be
constructed from the `free flows' given in (\ref{freesol}).
Take an arbitrary initial value $(q(0), p(0))$.
As a consequence of the Hamiltonian reduction, the corresponding
solution  $(q(t), p(t))$ of Hamilton's equation for $H_{k}^\red$
can be read off from the equality
\bea
&&\left(e^{q(t)}, J(q(t),p(t)),xC^l+ \xi(u^\kappa)\right) = \qquad
\label{sol}\\
&& \quad =\left(\eta(t)  e^{t V_{k}} e^{q(0)} h(t)^{-1},
\eta(t) J(q(0), p(0)) \eta(t)^{-1},
\eta(t)(xC^l+ \xi(u^\kappa)\eta(t)^{-1}\right),
\nonumber\eea
where
\be
V_{k}= J(q(0), p(0))^{2k-1} - \frac{1}{2n} \tr (J(q(0),p(0))^{2k-1}) \1_{2n}
\ee
and
$(\eta(t), h(t))\in G_+ \times G^+$ is determined by the condition that
the left-hand-side of (\ref{sol}) must belong to the gauge slice $S$ (\ref{S}).
Thus, finding the solution requires the determination of the generalized Cartan decomposition
\be
  e^{t V_{k}} e^{q(0)} = \eta(t)^{-1} e^{q(t)} h(t),
  \qquad
  (\eta(t), q(t), h(t))\in G_+ \times \cA_c \times G^+,
  \label{factor1}\ee
  made unique by the initial condition $\eta(0)=h(0)=\1_{2n}\in G$ and  continuity in $t$
  together with
  the auxiliary condition
  \be
  \eta(t) \xi(u^\kappa) \eta(t)^{-1} = \xi(u^\kappa).
  \ee
Then $p(t)$ obeys
  \be
  p(t)= \pi_-^-( \eta(t) J(q(0), p(0)) \eta(t)^{-1}) = \pi_-^-(\eta(t)L(q(0), p(0))\eta(t)^{-1}).
  \ee

If one is interested only in $q(t)$, then a simpler solution algorithm is also available.
For this, notice that
the evaluation of the expression $g \Gamma(g^{-1})$ (with $\Gamma$ defined by (\ref{2.7})) for both sides of
the equality in (\ref{factor1}) leads
to the relation
\be
e^{2 q(t)} D_m = \eta(t) e^{t V_k} e^{2q(0)} D_m
e^{t V_{k}^\dagger} \eta(t)^{-1}.
\label{factor2}\ee
This means that the entries of the diagonal matrix $e^{2 q(t)} D_m$ are the eigenvalues
of the Hermitian matrix
\be
e^{t V_{k}} e^{2q(0)} D_m
e^{t V_{k}^\dagger}.
\ee
It follows from (\ref{factor2}) and the forms of $D_m$ (\ref{2.3}) and the Weyl alcove (\ref{2.20}) that
the eigenvalues of the above Hermitian matrix are all different, and
therefore \emph{finding
$q(t)$ boils down to an ordinary diagonalization problem.}

The above algorithm could be particularly useful to analyze
the generalized Sutherland dynamics, which  arises as the $k=1$ special case of the reduced systems
(\ref{3.39}).
The formula (\ref{I1}) entails that in this case $p(t)= \dot{q}(t)$.
If desired,
one  could also use the above derivation to obtain a Lax representation for the
equations of motion, taking  $J(q,p)$ in (\ref{Jqp}),
or alternatively $L(q,p)$, as the Lax matrix.
However, the Lax pair would not give anything substantial  to our knowledge about the
generalized Sutherland model, whose
main features follow from its  realization as a reduction of the free geodesic motion on $SU(n,n)$.

\section{Conclusion}
\setcounter{equation}{0}

In this paper we applied the Hamiltonian reduction approach
to a particular  many-body system.
Our derivation of the generalized Sutherland model (\ref{I1}) by  reduction
of the free geodesic motion on the group $SU(n,n)$
proves the integrability of the model in the new case of three independent coupling
constants, and  provides a simple algorithm for constructing the solutions.
This potentially paves the way for future work to analyze the scattering
characteristics of the model
along the lines of the papers \cite{RIMS94,PG}.
The investigation  of the quantum mechanics of the model (for example by
quantum Hamiltonian reduction) is also a challenging problem.

Another interesting  problem is to find duality properties for the generalized Sutherland model,
which would extend the action-angle dualities of the integrable many-body systems
studied by Ruijsenaars (see e.g.~the review \cite{Banff}).
This problem exists in general for the Sutherland models with two types of particles,
whose duality properties are not even known in the $A_n$ case.
For the description of dualities in the reduction approach, see also \cite{FA,FK}
and references therein.

Recently \cite{Bog}
new integrable random matrix models have been constructed in association
with certain integrable many-body systems of Calogero-Sutherland type.
It could be feasible to extend this correspondence between
random matrix models and integrable-many body systems to other cases, possibly including
generalized Sutherland models with two types of particles.

We end with a remark on the Lax matrices that can be associated to the model (\ref{I1}).
Namely, we
note that our usage of $SU(n,n)$ as the starting point leads to a $2n \times 2n$
Lax matrix, but it should be also possible to derive a $(2n+1) \times (2n+1)$ Lax matrix
for the same model,
with 3 independent couplings, by reduction of the free motion
on $SU(n+1,n)$ (cf.~Ref.~\cite{FP1}).
In the case of equal couplings ($x = 2 y = 2 \sqrt{2}\kappa$ in (\ref{I1})), it is this latter Lax matrix
that one may expect to obtain directly as well from the standard Lax matrix of the
original Sutherland
model of $2n+1$ particles by applying imaginary shifts and restriction to `mirror symmetric'
configurations.
Although in the reduction approach the role of the Lax matrices is somewhat secondary, they
are central in other approaches \cite{Banff,Suth}. For this reason, we plan to describe
the  alternative Lax matrices of different size and their relationship elsewhere.

\bigskip
\medskip
\noindent{\bf Acknowledgements.}
This work was supported in part
by the Hungarian
Scientific Research Fund (OTKA) under the grant K 77400.
We thank J. Balog for useful discussions, and
thank the anonymous referee for pointing out an interesting question about alternative Lax matrices.

\end{document}